\RequirePackage{lineno}
\documentclass{emulateapj}

\usepackage{amsmath,natbib,graphicx}
\usepackage{epsf}
\usepackage{fancyvrb}
\usepackage{epstopdf}
\usepackage{epsfig}
\usepackage{color}
\DeclareGraphicsExtensions{.jpg,.pdf,.png,.eps,.ps}

\begin{document}

%\setpagewiselinenumbers
%\modulolinenumbers[1]
%\linenumbers

\VerbatimFootnotes
\title{STRONG DEPENDENCE OF THE INNER EDGE OF THE HABITABLE ZONE ON PLANETARY ROTATION RATE}

\shorttitle{Dependence of the HZ on Planetary Rotation Rate}

\author{Jun Yang}
\affil{Department of Geophysical Sciences, University of Chicago, Chicago, IL, USA}
\author{Gwena\"el Bou\'e}
\affil{Department of Astronomy \& Astrophysics, University of Chicago, Chicago, IL, USA and Sorbonne Universit\'es, UPMC Univ Paris 06, UMR 8028, IMCCE, Observatoire de Paris, F-75014 Paris, France}
\author{Daniel C. Fabrycky}
\affil{Department of Astronomy \& Astrophysics, University of Chicago, Chicago, IL, USA}
\and
\author{Dorian S. Abbot}
\affil{Department of the Geophysical Sciences, University of Chicago, Chicago, IL, USA}
\email{Correspondence: abbot@uchicago.edu}

\shortauthors{Yang, Bou\'e, Fabrycky, \& Abbot, ApJL, 2014}

\begin{abstract}
  Planetary rotation rate is a key parameter in determining
  atmospheric circulation and hence the spatial pattern of
  clouds. Since clouds can exert a dominant control on planetary
  radiation balance, rotation rate could be critical for determining
  mean planetary climate. Here we investigate this idea using a
  three-dimensional general circulation model with a sophisticated
  cloud scheme. We find that slowly rotating planets (like Venus) can
  maintain an Earth-like climate at nearly twice the stellar flux as
  rapidly rotating planets (like Earth). This suggests that many
  exoplanets previously believed to be too hot may actually be
  habitable, depending on their rotation rate.  The explanation for
  this behavior is that slowly rotating planets have a weak Coriolis
  force and long daytime illumination, which promotes strong
  convergence and convection in the substellar region. This produces a
  large area of optically thick clouds, which greatly increases the
  planetary albedo.  In contrast, on rapidly rotating planets a much
  narrower belt of clouds form in the deep tropics, leading to a
  relatively low albedo. A particularly striking example of the
  importance of rotation rate suggested by our simulations is that a
  planet with modern Earth's atmosphere, in Venus' orbit, and with
  modern Venus' (slow) rotation rate would be habitable.  This would
  imply that if Venus went through a runaway greenhouse, it had a
  higher rotation rate at that time. 
  
\end{abstract}


\section{Introduction}

It is traditional to define the habitable zone based on whether liquid
water can be maintained on a planet's surface, which is primarily
controlled by the planet's surface temperature \citep{Kastingetal1993,
  Kastingetal2014}. Accurate estimates of the stellar flux boundaries
of the habitable zone are critical for estimating parameters of
astrophysical interest such as the frequency of Earth-like planets
\citep[e.g.,][]{Kopparapu2013}.  The inner edge of the habitable zone
is set by the runaway greenhouse effect, a positive feedback through
which an entire ocean can be evaporated into the atmosphere
\citep{Nakajimaetal1992}. Our ability to constrain the stellar flux
corresponding to the inner edge of the habitable zone has been
severely hampered by the inability of 1D radiative-convective models
to predict cloud behavior \citep{Selsisetal2007}.

Clouds are critical to planetary energy balance.  Cloud reflection of
solar radiation is responsible for most of the planetary albedo on
modern Earth \citep{DonohoeandBattisti2011}, and clouds also
significantly increase Earth's greenhouse effect by absorbing
terrestrial infrared emission \citep{Harrisonetal1990}.  Cloud
coverage and location are primarily controlled by large-scale
atmospheric circulation, which is determined by a variety of factors
including stellar flux, orbital parameters, and rotation rate. As the
stellar flux increases, cloud coverage and thickness may increase,
potentially leading to a higher albedo and a negative feedback, or
decrease, potentially leading to a lower albedo and a positive
feedback\footnote{Here we assume that changes in cloud reflection of
  stellar radiation (cooling) dominate over changes in cloud
  absorption of planetary infrared radiation (warming), which is the
  case in the simulations we present below.}. Orbital parameters such
as obliquity and eccentricity can both drive large-amplitude seasonal
cycles of the atmospheric circulation and surface temperature
\citep[e.g.,][]{Ferreiraetal2014}, but they tend to minimally affect
the annual-mean climate \citep{WilliamsandPollard2002,
  WilliamsandPollard2003}.

%%%%%%%%%%%%%%%%%%%%%%%%%%%%%%%%%%%
\begin{figure*}[]
\begin{center}
\vspace{-22mm}
\begin{center}
\includegraphics[angle=0, width=34pc]{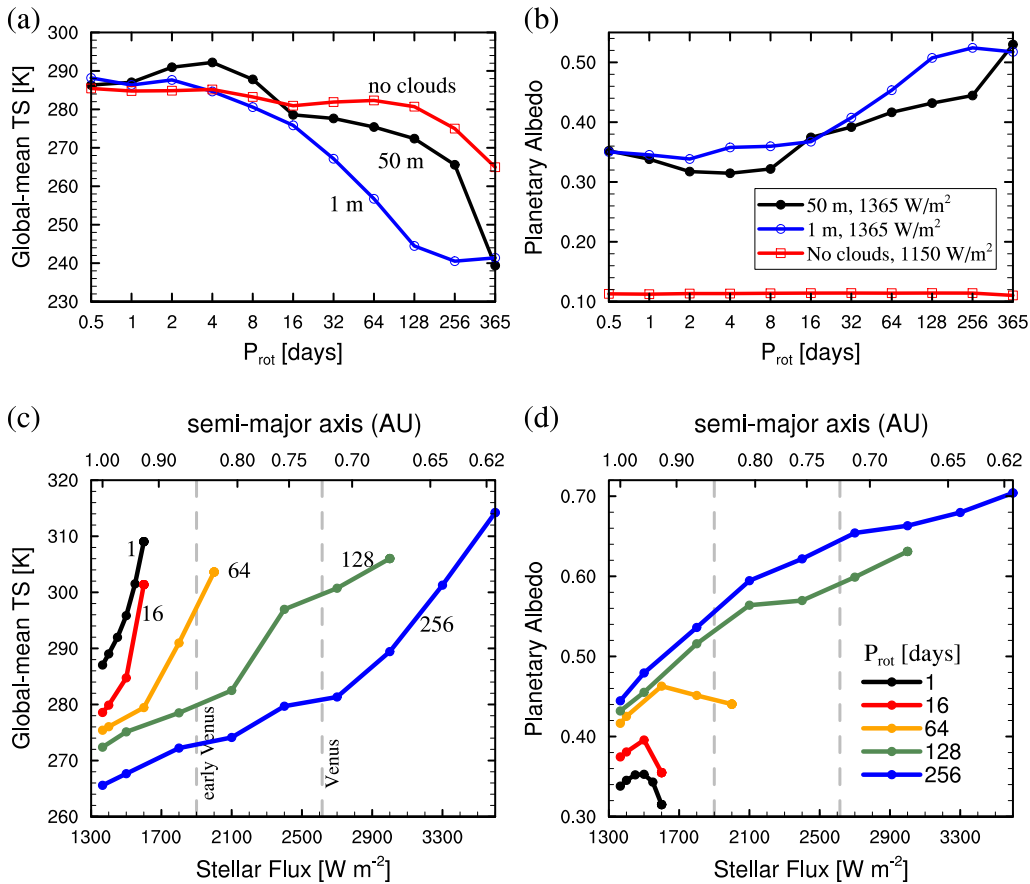}
\end{center}
\vspace{-8mm}
\caption{This figure demonstrates the dependence of planetary climate
  on rotation period ($P_{rot}$) for planets orbiting a Sun-like
  star. (a) and (b): Global-mean surface temperature (TS) and
  planetary albedo as a function of $P_{rot}$ for a given stellar flux
  ($S_0$). Black line: $S_0$\,=\,1365\,W\,m$^{-2}$ and the surface
  heat capacity (D) is equivalent to 50~m of water; blue line:
  $S_0$\,=\,1365\,W\,m$^{-2}$ and D\,=\,1~m; red line: clouds are
  switched off, $S_0$\,=\,1150\,W\,m$^{-2}$, and D\,=\,50~m.  For
  $P_{rot}$\,=\,365~days, the planet is in a synchronously rotating
  state. (c) and (d): Global-mean surface temperature (TS) and
  planetary albedo as a function of stellar flux ($S_0$) for a given
  $P_{rot}$ with D\,=\,50~m. The vertical dashed lines denote the
  stellar flux of early and modern Venus. The upper horizontal axis in
  (c--d) is the corresponding semi-major axis between a Sun-like star
  and the planet in AU.  In all these simulations the orbital period
  is 365 days and there is no sea ice.}
\label{fig1}
\end{center}
\end{figure*}

Planetary rotation rate determines the strength of the Coriolis force
(the apparent force felt due to the rotation of the planet) and the
length of day (and night). The Coriolis force is a key parameter in
determining the atmospheric circulation \citep[e.g.,][]{Schneider2006,
  Showmanetal2013}. If the Coriolis force is strong, thermally direct
latitudinal circulations (Hadley cells) are constrained to low
latitudes, and the atmosphere organizes into banded, roughly
longitudinally symmetric regions. If the Coriolis force is weak,
horizontal temperature gradients become small throughout the
atmosphere and the Hadley cells can extend globally.  The length of
day, combined with surface and atmospheric thermal inertia, helps
determine the surface temperature distribution, which drives
atmospheric circulation \citep{Pierrehumbert2010}. For a short day (or
large thermal inertia), the surface temperature difference between day
and night is small. If the day is long enough that the dayside is much
warmer than the nightside, atmospheric circulation is characterized by
ascent on the warm dayside and descent on the cold nightside.

Recently a number of calculations have been done with 3D general
circulation models (GCMs) to assess the effects of atmospheric
circulation, subsaturation, and clouds on the inner edge of the
habitable zone
\citep{Yangetal2013,Leconteetal2013b,WolfandToon2014}. In the extreme
case of tidally locked synchronously rotating planets orbiting
M-stars, strong atmospheric ascent on the dayside leads to thick
dayside cloud coverage and a very high planetary albedo
\citep{Yangetal2013}. This can allow a planet to remain habitable at
twice the stellar flux 1D model calculations would suggest.  In
contrast, for rapidly rotating planets with banded atmospheric
circulations, cloud behavior remains roughly similar to modern Earth's
(which the albedo of 1D models are tuned to) so that the inner edge of
the habitable zone in 3D models is similar to that in 1D models
\citep{Leconteetal2013b,WolfandToon2014}. Another interesting
difference is that the cloud feedback near the runaway greenhouse
threshold appears to be negative for tidally locked planets
\citep{Yangetal2013} and positive for rapidly rotating planets
\citep{Leconteetal2013b,WolfandToon2014}\footnote{\citet{WolfandToon2014}
  find a cloud feedback that starts negative, then becomes positive
  near the runaway greenhouse.}.

%%%%%%%%%%%%%%%%%%%%%%%%%%%%%%%%%%%
%%%%%%%%%%%%%%%%%%%%%%%%%%%%%%%%%%

\begin{figure*}[]
%\begin{center}
\vspace{-10mm}
\begin{center}
\includegraphics[angle=0, width=34pc]{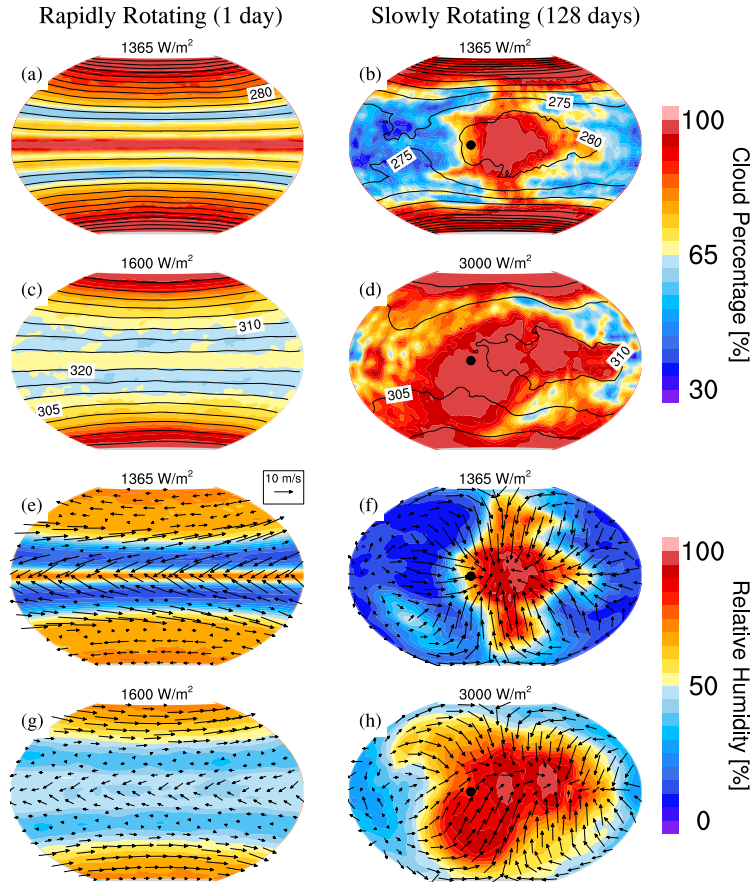}
\end{center}
\vspace{-22mm}
\caption{Differences in clouds and atmospheric circulation between
  rapidly (left) and slowly (right) rotating planets with an orbital
  period of 365~days. The rotation period is 1~day for the rapidly
  rotating planet and the stellar flux is
  1365~or~1600~W\,m$^{-2}$. The rotation period is 128~days for the
  slowly rotating planet and the stellar flux is
  1365~or~3000~W\,m$^{-2}$.  (a--d): Total cloud coverage (\%, shaded)
  and surface temperature (K, black contours with an interval of
  5~K). (e--h): Relative humidity at 450 mbar (\%, shaded)
  and near-surface winds (m\,s$^{-1}$, vectors).  The black dot in the
  right panels is the transient substellar point, which moves
  westward around the planet with a period of 197~days. All variables
  are averaged over 30 days.}
\label{fig2}
%\end{center}
\end{figure*}

The purpose of this study is to investigate the effects of a range of
planetary rotation rates, between tidally locked and rapidly rotating,
and stellar types on cloud behavior and the inner edge of the
habitable zone. To do this we use a 3D GCM with a sophisticated
cloud scheme that reproduces cloud behavior well in the large range of
climates observed on modern Earth. Although we do not push the model
significantly outside of this range, it is important to note that
cloud modeling is difficult, and other models may yield quantitatively
different results. Nevertheless, we focus on results due to
robust physical processes that should be qualitatively similar in any
3D model. Our main conclusion is that for all stellar types slowly
rotating planets (orbital period $\approx$100 days or more) behave
similarly to tidally locked planets and have a high planetary albedo
near the inner edge of the habitable zone. The width of the habitable
zone is therefore strongly dependent on planetary rotation rate.

%%%%%%%%%%%%%%%%%%%%%%%%%%%%%%%%%%%
%%%%%%%%%%%%%%%%%%%%%%%%%%%%%%%%%%%

\section{METHODS}

For most of the results presented below we use the Community
Atmosphere Model version 3.1 \citep[CAM3,] []{Collinsetal2004}, which
is a 3D atmospheric general circulation model (GCM) that was developed
by the National Center for Atmospheric Research to simulate the
climate of Earth. CAM3 calculates atmospheric circulation and
radiative transfer, and uses subgrid-scale parameterizations to model
small-scale vertical convection, clouds, and precipitation. We perform
additional simulations in some cases using the Community Atmosphere
Model version 4.0 \citep[CAM4,][]{Nealeetal2010} and the Community
Climate System Model version 3.0
\citep[CCSM3,][]{Collinsetal2006}. CCSM3 is a coupled ocean-atmosphere
model that calculates ocean circulation explicitly and uses CAM3 as
its atmospheric component. We run CAM3 and CAM4 coupled to an immobile
ocean with a uniform depth and a uniform albedo of 0.06.  The GCMs
simulate marine stratus, layered, shallow convective, and deep
convective clouds as well as prognostically calculate liquid and ice
cloud condensate.  Compared to CAM3, CAM4 has a revised deep
convection scheme and a similar cloud scheme.
The default atmospheric pressure we use is 1.0 bar (mainly N$_2$), with
CO$_2$=400~ppmv and CH$_4$=1~ppmv. We set geothermal heat flux,
aerosols, obliquity, and eccentricity to zero. We run CAM3 at a
horizontal resolution of 3.75$^{\circ}$$\times$3.75$^{\circ}$ with 26
vertical levels from the surface to $\approx$30 km.

\begin{figure*}[]
%\begin{center}
\vspace{-16mm}
\begin{center}
\includegraphics[angle=0, width=34pc]{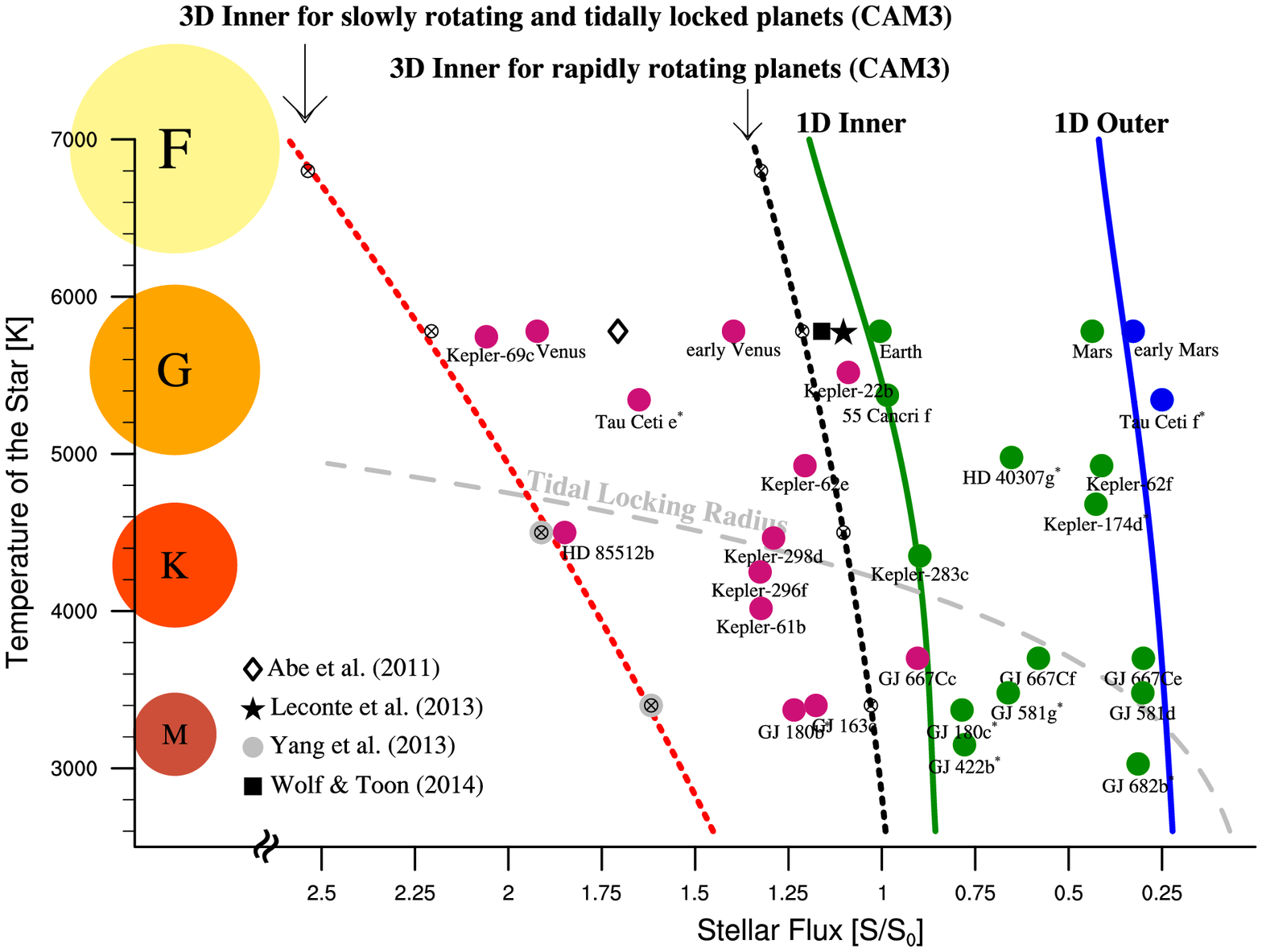}
\end{center}
\vspace{-28mm}
\caption{Habitable zone boundaries as a function of stellar type and
  planetary rotation rate for a 1D radiative-convective model and for
  the 3D general circulation model CAM3. Blue line: the 1D outer edge
  \citep[maximum greenhouse,][]{Kopparapuetal2013}; green line: the 1D
  inner edge \citep[runaway greenhouse,][]{Kopparapuetal2013}; black
  line: the 3D inner edge for rapidly rotating planets in CAM3
  (rotation period of 1~day); red line: the 3D inner edge for slowly
  rotating planets in CAM3 (rotation period of 128~days for G and F
  stars, and tidally locked with an orbit of 60~days for M and K
  stars); gray line: the tidal locking radius \citep{Kastingetal1993}.
  The CAM3 simulations used to calculate the 3D inner edge lines are
  denoted by $\otimes$. We also plot the inner edge of the habitable
  zone for rapidly rotating dry planets \citep{Abeetal2011}, for Earth
  obtained in generic-LMD \citep{Leconteetal2013b} and CAM3
  with a modified radiative-transfer module \citep{WolfandToon2014}. 
  Finally, we plot solar system planets and
  discovered exoplanets (unconfirmed exoplanets are marked by
  $^{\ast}$).}
\label{fig3}
%\end{center}
\end{figure*}

In our simulations where we change the rotation rate and increase the
stellar flux we use Earth's planetary parameters, including radius
($R_\oplus$), gravity ($g_\oplus$), and orbital period ($P_{orb}$, 365
days). We increase the stellar flux until the model crashes, which
occurs when the global mean surface temperature reaches
$\approx$310~K. Comparison with the clear-sky calculations of
\citet{Leconteetal2013b} shows agreement between CAM3 and generic-LMD
to within $\approx$5~K up to this temperature. We use the last
converged solution as a proxy for the inner edge of the habitable
zone. We examine a series of rotation periods ($P_{rot}$) ranging from
12 hours to 365 days.  We perform simulations using the Sun's spectrum
and stellar spectra corresponding to M (blackbody 3400~K), K (4500~K),
and F (6800 K) stars. In our simulations examining a planet with
Venus' orbital characteristics and Earth's atmosphere, we use a
stellar flux $S_0$\,=\,2615\,W\,m$^{-2}$, $P_{orb}$\,=\,225 days,
$P_{rot}$\,=\,$-$243 days (retrograde rotation),
$R_p$\,=\,0.95\,$R_\oplus$, and $g_p$\,=\,0.9\,$g_\oplus$.  Our
default simulation in this case uses Earth's continental configuration
with continents composed of clay and sand (albedo=0.2, thermal
conductivity=0.26 W\,m$^{-1}$\,K$^{-1}$).

%%%%%%%%%%%%%%%%%%%%%%%%%%%%%%%%%%%
%%%%%%%%%%%%%%%%%%%%%%%%%%%%%%%%%%%

\section{DRAMATIC EFFECT OF PLANETARY ROTATION}

For a given stellar flux ($S_0$), the surface temperature (TS) of
rapidly rotating planets is much higher than that of slowly rotating
planets (Fig.~1a). When we use a surface heat capacity equivalent to
50~m of water (D=50~m), the global-mean TS decreases by 20~K when we
increase the rotation period ($P_{rot}$) from 1~day to 256~days, and
then decreases by another 25~K when the planet becomes tidally locked
($P_{rot}$=365~days). The primary reason for this is that the cloud
albedo is much higher on slowly rotating planets (Fig.~1b). If we
artificially set clouds to zero in the model, TS is nearly independent
of rotation rate, except for the tidally locked and nearly tidally
locked cases (Fig.~1a).  TS is lower in these cases because a thermal
inversion on the nighside leads to a radiator fin that cools the
planet \citep{Yangetal2013, YangandAbbot2014}. If we reduce D to 1~m,
the TS distribution is able to adjust to the moving stellar forcing so
that the planet can behave as if it were tidally locked even when
$P_{rot}$ is decreased to 128~days (Fig.~1). When D=50~m, TS has a
maximum at $P_{rot}$\,=\,4 days that appears to be associated with a
high latitude cloud feedback similar to that described by
\citet{AbbotandTziperman2008}.

\begin{figure*}[]
%\begin{center}
\vspace{-30mm}
\begin{center}
\includegraphics[angle=0, width=34pc]{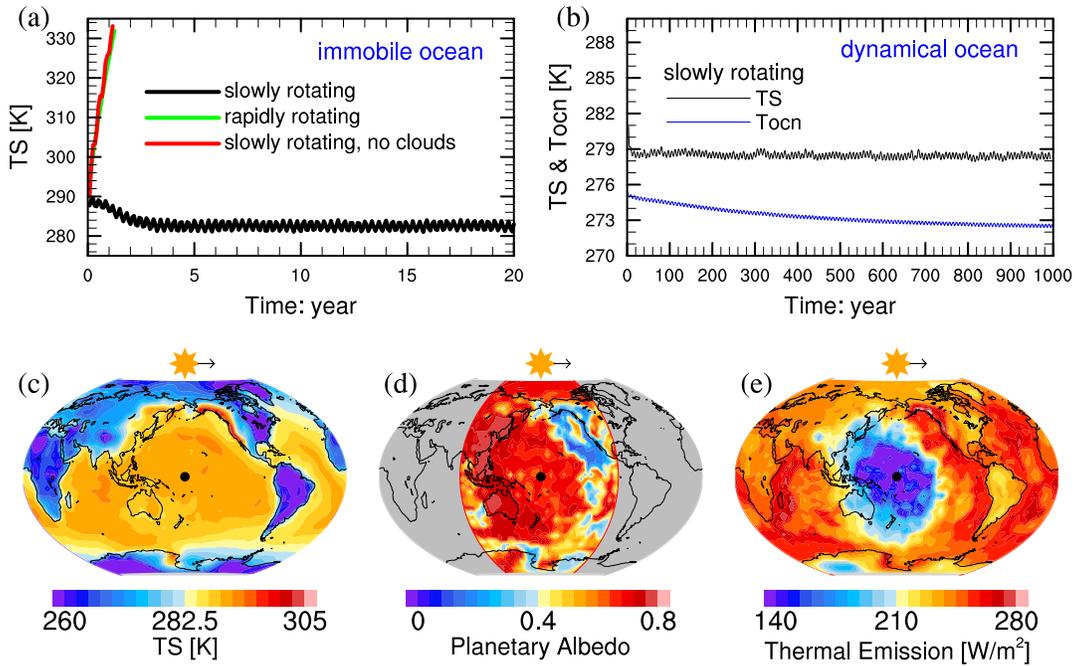}
\end{center}
\vspace{-30mm}
\caption{The climate of a planet with modern Earth's atmosphere and
  continental configuration, but in Venus' orbit and with Venus'
  (slow) rotation rate. (a): Time series of global-mean surface
  temperature (TS) simulated by the atmospheric model CAM3 for Venus'
  rotation rate (slowly rotating, black), Earth's rotation rate
  (rapidly rotating, green), and Venus' rotation rate with clouds
  artificially set to zero (slowly rotating, no clouds, red). The
  planet quickly tends toward a runaway greenhouse if it is rapidly
  rotating or has no clouds, but is habitable if it is slowly
  rotating. (b): Global-mean TS and
  vertically-integrated ocean temperature (Tocn) in a coupled
  ocean-atmosphere simulation using CCSM3 and Venus' rotation
  rate. (c--e): maps of TS, planetary albedo, and
  thermal emission to space averaged over 1~day in the coupled CCSM3
  simulation.  The black dot in (c--e) is the transient substellar
  point, which moves eastward around the planet with a period of 117
  days.}
\label{fig4}
%\end{center}
\end{figure*}

There is a dramatic difference between the response of the planetary
albedo to increases in $S_0$ for rapidly and slowly rotating
planets. For rapidly rotating planets, as $S_0$ increases the
planetary albedo first increases, then decreases (Fig.~1d), leading to
a positive feedback near the runaway greenhouse threshold
\citep{Leconteetal2013b,WolfandToon2014}. This positive feedback is
due to a combination of decreased cloud reflection and increased water
vapor absorption. In contrast, for slowly rotating planets, the
planetary albedo monotonously increases with $S_0$, leading to a
negative feedback that stabilizes the climate. For rapidly rotating
planets, the atmospheric circulation is banded and Earth-like. This
leads to high cloud coverage both in a tropical convergence zone
associated with the ascent of the Hadley cells (Fig.~2e) and at higher
latitudes associated with baroclinic eddies (Fig.~2a). The tropical
clouds are most important for planetary albedo because the stellar
flux is highest there. As $S_0$ increases, the equator-to-polar
temperature gradient decreases (Fig.~2c), which weakens the Hadley
cells \citep{HeldandHou1980}, reduces tropical cloud coverage
(Fig.~2c), and decreases the planetary albedo (Fig.~1d). For slowly
rotating planets, a global atmospheric circulation occurs with strong
low-level convergence and ascent in the (slowly moving) substellar
region (Fig.~2f) similar to the circulation on synchronously rotating
planets \citep{Joshietal1997,Showmanetal2013,Leconteetal2013a}. This
circulation leads to strong convection and optically thick clouds
(Fig.~2b) in the substellar region \citep{Yangetal2013}. As $S_0$
increases the circulation weakens, but the zone of ascent spreads out,
which leads to a broader area of high relative humidity (Fig.~2h) and
high cloud coverage (Fig.~2d).  Additionally, the cloud water content
increases, making individual clouds optically thicker. As a result,
the planetary albedo increases with $S_0$ (Fig.~1d).

Our simulations indicate that the inner edge of the habitable zone is
strongly dependent on planetary rotation rate. Numerous discovered
exoplanets that were previously considered uninhabitable may be within
the habitable zone\footnote{Our simulations generally have low
  stratospheric water vapor (Table~1), but a full investigation of
  water loss on slowly rotating planets is beyond the scope of this
  paper.} if they rotate slowly (Fig.~3). Our simulations yield an
inner edge of the habitable zone for rapidly rotating planets
approximated by the curve
\begin{equation}
\begin{split}
S_{rap}=1.2138+9.8344\times10^{-5}\,(T_{eff}-5780)+\\
8.8000\times10^{-9}\,(T_{eff}-5780)^{2},
\end{split}
\end{equation}
%
%\[S_{rap}=1.2138+9.8344\times10^{-5}\,(T_{eff}-5780)+\\
%8.8000\times10^{-9}\,(T_{eff}-5780)^{2},\]
%
and for slowly rotating planets 
\begin{equation}
\begin{split}
S_{slow}=2.2296 + 2.8056\times10^{-4}\,(T_{eff}-5780)+\\
1.1308\times10^{-8}\,(T_{eff}-5780)^{2},
\end{split}
\end{equation}

%\[S_{slow}=2.2296 + 2.8056\times10^{-4}\,(T_{eff}-5780)+
%1.1308\times10^{-8}\,(T_{eff}-5780)^{2},\]
%
where $S_{rap}$ and $S_{slow}$ are stellar fluxes divided by
1360~W\,m$^{-2}$ and $T_{eff}$ is the temperature of the star. We note
that our $S_{rap}$ is slightly larger than that of \citet{WolfandToon2014},
who also use CAM3, but have implemented a more sophisticated radiative
transfer scheme, and use different surface boundary conditions,
orbital parameters, and trace gas concentrations (Fig.~3). Although
the values of $S_{rap}$ and $S_{slow}$ estimated using different GCMs
and assumptions will likely differ by $\approx$10\%, the robust cloud
processes we describe should always make $S_{slow}$ significantly
larger than $S_{rap}$.

%\begin{table}[t]
\begin{table*}
  \vspace{0mm}
  Table~1: Simulated climates of a planet with Venus' orbit and (slow) rotation rate 
  and with an Earth-like atmosphere. By default, S$_0$\,=\,2615\,W\,m$^{-2}$, 
  $P_{rot}$\,=\,$-$243~days, and $P_{orb}$\,=\,225~days.
  \vspace{0mm}
  \begin{center}
\begin{tabular}{lllccccc}
  \hline
  Group & Model & Experimental Design  & $\alpha_p$$^{a}$ & G$^{b}$ & V(H$_2$0)$^{c}$ & TS$^{d}$    \\
  &            &                               & 0--1       &  K  & ppmv & K                   \\
  \hline
  1$^{e}$ & CAM3   & default (Earth's continental configuration)  & 0.64 & 30 & 1.8 & 282 \\
  1 & CAM3   & no continents (aqua-planet)                                  & 0.58 & 39 & 6.5 & 303\\
  1 & CAM4   & no continents (aqua-planet)                                  & 0.59 & 38 & 49 & 300\\
1 & CCSM3 & dynamical ocean              & 0.64  & 24 & 152 & 278 \\
1 & CCSM3 & dynamical ocean (4~km), no continents         & 0.64  & 29 & 0.4 & 283 \\
\hline
2$^{f}$ & CAM3   & default (Earth's continental configuration) & 0.65 & 31 & 2.0 & 284 \\
%2 & CAM3   & with vegetation, lakes, and polar ice sheets  & 0.65 & 31 & 1.7 & 283\\
2 & CAM3   & switch off sea ice module                        & 0.65 & 31 & 2.0 & 284\\ 
2 & CAM3   & no continents (aqua-planet)                   & 0.59 & 41 & 7.5 & 303\\
2 & CAM3   & no continents, increase the model resolution   & 0.62 & 46 & 137 & 307\\
2 & CAM3   & decrease the model time step                           & 0.64 & 32 & 2.1 & 284\\
2 & CAM3   & cloud ice particle size\,$\times$\,0.5        & 0.66 & 32 & 10 & 282 \\
2 & CAM3   & cloud ice particle size\,$\times$\,2.0           & 0.63 & 31 & 0.6 & 285\\
2 & CAM3   & cloud liquid particle size\,$\times$\,0.5    & 0.67 & 29 & 0.5 & 276\\
2 & CAM3   & cloud liquid particle size\,$\times$\,2.0       & 0.58 & 43 & 454 & 306\\
2 & CAM3   & decrease the mixed layer depth to 1~m         & 0.64 & 26 & 4.1 & 280\\
2 & CAM3   & increase the mixed layer depth to 200~m       & 0.65 & 31 & 1.6 & 283\\
2 & CAM3   & increase the mixed layer depth to 500~m       & 0.65 & 32 & 1.7 & 284\\
2 & CAM3   & increase the CO$_2$ concentration to 0.1 bar & 0.63 & 56 & 158 & 310\\
2 & CAM3   & increase the surface pressure to 2 bars   & 0.65  & 36 & 0.3 & 287\\
2 & CAM3   & increase the surface pressure to 5 bars    & 0.65 & 49 & 0.3 & 301 \\
2 & CAM3   & increase gravity to 1.5\,$g_\oplus$     & 0.64 & 29 & 0.7 & 283\\
2 & CAM3   & increase the radius to 2.0\,$R_\oplus$      & 0.65 & 31 & 2.0 &  284 \\
2 & CAM3   & increase the obliquity from 0$^{\circ}$ to 23.5$^{\circ}$  & 0.65 & 31 & 1.9 & 283\\
2 & CAM3   & increase the eccentricity from 0 to 0.2                & 0.61 & 41 & 611 & 301\\
\hline
3$^{g}$ & CAM3   & S$_0$\,=\,1900\,W\,m$^{-2}$, $P_{rot}$\,=\,8~days  &  \multicolumn{4}{c}{tends to runaway warming} \\
3 & CAM3               & S$_0$\,=\,1900\,W\,m$^{-2}$, $P_{rot}$\,=\,16~days  & 0.46 & 41 & 0.6 &  301\\
3 & CAM3               & S$_0$\,=\,2615\,W\,m$^{-2}$, $P_{rot}$\,=\,32~days  &  \multicolumn{4}{c}{tends to runaway warming}  \\
3 & CAM3               & S$_0$\,=\,2615\,W\,m$^{-2}$, $P_{rot}$\,=\,48~days  & 0.63 &  32 & 2.4 & 287 \\
\hline
\end{tabular}
 \end{center} 
 \vspace{-3mm}
 \emph{a.}~$\alpha_p$: planetary albedo. \\
 \emph{b.}~G: global-mean greenhouse effect. \\
 \emph{c.}~V(H$_2$0): stratospheric vapor content in units of parts per million by volume (ppmv). \\
 \emph{d.}~TS: global-mean surface temperature.\\
 \emph{e.}~Group 1: pCO$_2$\,=\,400 ppmv, pCH$_4$\,=\,1 ppmv, pN$_2$O\,=\,0.\\
 \emph{f.}~Group 2: pCO$_2$\,=\,500 ppmv, pCH$_4$\,=\,10 ppmv, pN$_2$O\,=\,1 ppmv.\\
 \emph{g.}~Group 3: same as default sets of Group~1, except varying S$_0$ and/or $P_{rot}$. \\

\end{table*}

Our results suggest that a planet at Venus' distance from a Sun-like
star could be habitable if it rotated slowly and had an Earth-like
atmosphere\footnote{We note that
  \citet{Leconteetal2013b} speculate that rotation rate could be
  important for the history of habitability on Venus.}. We checked
this using CCSM3 to simulate the climate a planet like Earth would
have with Venus' present orbital parameters and rotation rate
($P_{rot}$\,=\,$-$243 days). Although the stellar flux is 1.92 times
modern Earth's, the planetary albedo is 0.65 so that the planet
absorbs slightly less radiation than modern Earth and the maximum TS
is only 306~K (Fig.~4).  The slow rotation and consequent stabilizing
cloud feedback are the keys to preventing the planet from entering a
runaway greenhouse at this high stellar flux. If either $P_{rot}$ is
decreased to 1 day or the clouds are switched off, the model tends
toward a runaway greenhouse (Fig.~4a).  Further simulations show that
the planet tends toward a runaway greenhouse if either the day length
or the Coriolis parameter are changed to be Earth-like (and the other
is held constant). We have performed a large variety of sensitivity
tests that demonstrate the robustness of our conclusion that an
Earth-like planet in Venus' orbit would likely be habitable
(Table~1). These ideas could eventually be tested by using the James
Webb Space Telescope to look for weak thermal emission at the
substellar point (Fig.~4e) of detected exoplanets
\citep{Yangetal2013}.

Our work has important implications for the evolution of
Venus. Deuterium enrichment in Venus' atmosphere suggests that it may
have started with an ocean and gone through a runaway greenhouse
\citep{Donahueetal1982}. Our results suggest that if the runaway
happened near the beginning of the solar system, Venus would have had
to have a rotation period less than a few weeks, and if the runaway
occurred recently, Venus would have had to have a rotation period less
than a few months (Fig.~1, Table~1). Any water would then be
photodissociated and lost to space, and large amounts of CO$_2$ would
accumulate in the atmosphere since silicate weathering would not occur
if there were no surface water \citep{Kasting2010}.  Tidal
interactions could later slow Venus' rotation rate to its present
value \citep{CorreiaandLaskar2001}, but it would be too late for Venus
to return to habitable conditions. If instead all of the water on
Venus were lost through hydrodynamic escape during formation and Venus
did not start with oceans \citep{Hamanoetal2013}, then the rotation
rate would not need to have changed over its history.

Consideration of  Venus shows that  slowly rotating planets  which our
calculations suggest \textit{could} be  habitable will not actually be
habitable  in \textit{all}  cases. This  will depend  on  whether they
started with a  rapid rotation rate, if so how long  it took for their
rotation  rate to  slow, the  rate  of water  loss if  they entered  a
runaway greenhouse  at some point  in their history,  their (possible)
migration  history, and the  timing and  amount of  volatile delivery,
among other things. This is no different from any previous estimate of
habitable zone, since it  has always been understood that habitability
depends on planetary history in addition to location.

%%%%%%%%%%%%%%%%%%%%%%%%%%%%%%%%%%%
%%%%%%%%%%%%%%%%%%%%%%%%%%%%%%%%%%%

\section{Conclusions}

This work demonstrates that the inner edge of the habitable zone for
slowly rotating planets could be at twice the stellar flux as for
rapidly rotating planets. Numerical simulations suggest that the
rotation periods of planets at formation could vary between 10 hours
and 400 days \citep{MiguelandBrunini2010}, and tidal interactions can
further slow planetary rotation \citep{LagoandCazenave1979}. It is
therefore probable that a large number of planets rotate slowly enough
to have a greatly expanded habitable zone. Additionally, our
simulations suggesting that an Earth-like planet with Venus' present
orbit and rotation rate would be habitable demonstrate that empirical
limits on the habitable zone based on solar system planets
\citep[e.g.,][]{Kastingetal2014} may not be as strong constraints as
previously believed, depending on factors such as rotation rate and
planetary history. Finally, we note that although we expect our
results to be qualitatively robust, the details will differ with other
models that have different cloud and radiation schemes. We can hope to
resolve this issue by comparing GCMs, applying cloud resolving models
to the problem \citep[e.g.,][]{Abbot2014}, and eventually observing
planets using methods such as those suggested by \citet{Yangetal2013}.

%%%%%%%%%%%%%%%%%%%%%%%%%%%%%%%%%%%
%%%%%%%%%%%%%%%%%%%%%%%%%%%%%%%%%%%

\acknowledgments \textbf{Acknowledgments:} We are grateful to D.~Koll, Y.~Wang, Y.~Liu, 
F. Ding, C. Zhou, and C.~Bitz for technical assistance and/or helpful discussions. 
  D.S.A. acknowledges support from an Alfred P. Sloan Research Fellowship. 
  This work was completed in part with resources provided by 
  the University of Chicago Research Computing Center. 
 %%%%%%%%%%%%%%%%%%%%%%%%%%%%%%%%%%%
%%%%%%%%%%%%%%%%%%%%%%%%%%%%%%%%%%%

%\clearpage


\begin{thebibliography}{}
\bibitem[Abbot(2014)]{Abbot2014} Abbot, D. S. 2014, J. Climate, in press
\bibitem[Abbot \& Tziperman(2008)]{AbbotandTziperman2008} Abbot, D. S., \& Tziperman, E. 2008, QJRMS, 134, 165
\bibitem[Abe et al.(2011)]{Abeetal2011} Abe, Y., Abe-Ouchi, A., Sleep, N. H., \& Zahnle, K. J.,  2011, Astrobiology, 11, 443
\bibitem[Collins et al.(2004)]{Collinsetal2004} Collins, W. D., Basch, P. J., Boville, B. A., et al. 2004, Technical Note, Document NCAR-TN-464+STR (Boulder, USA: NCAR)
\bibitem[Collins et~al.(2006)]{Collinsetal2006} Collins, W. D., Bitz, C. M., Blackmon, M. L., et al. 2006, J. Climate, 19, 2122
\bibitem[Correia \& Laskar(2001)]{CorreiaandLaskar2001} Correia, A. C. M., \& Laskar, J. 2001, Nature, 411, 767
\bibitem[Donahue et al.(1982)]{Donahueetal1982} Donahue, T. M., Hoffman, J. H., Hodges, R. R., \& Watson, A. J. 1982, Science, 216, 630
\bibitem[Donohoe \& Battisti(2011)]{DonohoeandBattisti2011} Donohoe, A., \& Battisti, D. S. 2011, J. Climate, 24, 4402
\bibitem[Ferreira et al.(2014)]{Ferreiraetal2014} Ferreira, D., Marshall, J., O'Gorman, P. A., \& Seager, S. 2014, Icarus, submitted
\bibitem[Hamano et al.(2013)]{Hamanoetal2013} Hamano, K., Abe, Y., \& Genda, H. 2013, Nature, 497, 607
\bibitem[Harrison et al.(1990)]{Harrisonetal1990} Harrison, E. F.,
  Minnis, P., Barkstrom, B. R., Ramanathan, V., Cess, R. D., \&
  Gibson, G. G. 1990, J. Geophys. Res, 95, 18,687
\bibitem[Held \& Hou(1980)]{HeldandHou1980} Held, I. M., \& Hou, A. Y. 1980, J. Atmos. Sci., 37, 515
\bibitem[Joshi et al.(1997)]{Joshietal1997} Joshi, M. M., Haberle, R. M., \& Reynolds, R. T. 1997, Icarus, 129, 450 
\bibitem[Kasting et al.(1993)]{Kastingetal1993} Kasting, J. F., Whitmire, D. P., \& Reynolds, R. T. 1993, Icarus, 101, 108
 \bibitem[Kasting(2010)]{Kasting2010} Kasting, J. F. 2010, How to find a habitable planet (Princeton, USA: Princeton University Press)
\bibitem[Kasting et al.(2014)]{Kastingetal2014} Kasting, J. F., Kopparapu, R., Ramirez, R. M., \& Harman, C. E. 2014, PNAS, in press
\bibitem[Kopparapu (2013)]{Kopparapu2013} Kopparapu, R. K. 2013, ApJL, 767 L8
\bibitem[Kopparapu et al.(2013)]{Kopparapuetal2013} Kopparapu, R. K., Ramirez, R., Kasting, J. F., et al. 2013, \apj, 767, 131
\bibitem[Lago \& Cazenave(1979)]{LagoandCazenave1979} Lago B., \& Cazenave A. 1979, Moon and Planets, 21, 127
\bibitem[Leconte et al.(2013a)]{Leconteetal2013a} Leconte, J., Forget, F., Charnay, B., et al. 2013a, A\&A, 554, A69
\bibitem[Leconte et al.(2013b)]{Leconteetal2013b} Leconte, J., Forget, F., Charnay, B., Wordsworth, R., \& Pottier, A. 2013b, Nature, 504, 268
\bibitem[Miguel \& Brunini(2010)]{MiguelandBrunini2010} Miguel, Y., \& Brunini, A., 2010, MNRAS, 406, 1935
\bibitem[Nakajima et al.(1992)]{Nakajimaetal1992} Nakajima, S.,
  Hayashi, Y.-Y., \& Abe, Y. 1992, J. Atmos. Sci., 49, 2256
\bibitem[Neale et al.(2010)]{Nealeetal2010} Neale, R. B., Richter, J. H., Conley, A. J., et al. 2010, Technical Note, Document NCAR-TN-485+STR (Boulder, USA: NCAR)
 \bibitem[Pierrehumbert(2010)]{Pierrehumbert2010} Pierrehumbert, R. T. 2010, Principles of Planetary Climate (Cambridge, UK: Cambridge University Press)
 \bibitem[Schneider(2006)]{Schneider2006} Schneider, T. 2006, Annu. Rev. Earth Planet. Sci., 34, 655
\bibitem[Selsis et al.(2007)]{Selsisetal2007} Selsis, F., Kasting, J. F., Levrard, B., et al. 2007, A\&A, 476, 1373
\bibitem[Showman et al.(2013)]{Showmanetal2013} Showman, A. P., Wordsworth, R. D., Merlis, T. M., \& Kaspi, Y. 2013, in Comparative Climatology of Terrestrial Planets, eds. Mackwell, S. J., et al.,  University of Arizona Press, Arizona, USA, pp. 277-326.
\bibitem[Williams \& Pollard(2002)]{WilliamsandPollard2002} Williams, D. M., \& Pollard, D. 2002, International Journal of Astrobiology, 1, 61
\bibitem[Williams \& Pollard(2003)]{WilliamsandPollard2003} Williams, D. M., \& Pollard, D. 2003, International Journal of Astrobiology, 2, 1
\bibitem[Wolf \& Toon(2014)]{WolfandToon2014} Wolf, E. T., \& Toon, O. B. 2014, Geophys. Res. Lett., 41, 167
\bibitem[Yang et al.(2013)]{Yangetal2013} Yang, J., Cowan, N. B., \& Abbot, D. S. 2013, ApJL, 771, L45
\bibitem[Yang \& Abbot(2014)]{YangandAbbot2014} Yang, J., \& Abbot, D. S. 2014, \apj, 784, 155

\end{thebibliography}
\end{document}